\documentclass[twocolumn]{aastex61}
\usepackage{newtxtext,newtxmath}
\usepackage[T1]{fontenc}
\usepackage{ae,aecompl}

\usepackage[colorinlistoftodos]{todonotes}
\usepackage{graphicx}	
\usepackage{amsmath}	
\usepackage{amssymb}	
\usepackage{aas_macros}
\usepackage{longtable}
\usepackage{color}
\usepackage{multirow}
\usepackage[english]{babel}

\newcommand{\sgra}{Sgr~A$^*$}
\newcommand{\magnt}{PSR~J1745$-$2900}

\newcommand{\lf}{$L_\text{f}$}
\newcommand{\parallellos}{\mathbin{\!/\mkern-5mu/\!}}

\received{}
\revised{}
\accepted{}
\submitjournal{ApJ}

\shorttitle{Large magnetic variations towards the Galactic Centre magnetar}
\shortauthors{Desvignes et al.}

\begin{document}

\title{Large magnetic field variations towards the Galactic Centre magnetar, PSR~J1745$-$2900}

\correspondingauthor{Gregory Desvignes, Ralph P. Eatough}
\email{gdesvignes@mpifr-bonn.mpg.de, reatough@mpifr-bonn.mpg.de}

\author{G. Desvignes}
\affil{Max-Planck-Institut f\"ur Radioastronomie, Auf dem H\"ugel 69, D-53121 Bonn, Germany}

\author{R. P. Eatough}
\affil{Max-Planck-Institut f\"ur Radioastronomie, Auf dem H\"ugel 69, D-53121 Bonn, Germany}

\author{U. L. Pen}
\affil{Canadian Institute for Theoretical Astrophysics, University of Toronto, 60 St. George Street, Toronto, Ontario M5S 3H8, Canada}
\affil{Dunlap Institute for Astronomy and Astrophysics, University of Toronto, 50 St. George Street, Toronto, Ontario M5S 3H4, Canada}
\affil{Canadian Institute for Advanced Research, CIFAR Program in Gravitation and Cosmology, Toronto, Ontario M5G 1Z8, Canada}
\affil{Perimeter Institute for Theoretical Physics, 31 Caroline Street North, Waterloo, Ontario, N2L 2Y5, Canada}

\author{K.~J.~Lee}
\affil{Kavli institute for astronomy and astrophysics, Peking University, Beijing 100871, P.R. China}

\author{S.~A.~Mao}
\affil{Max-Planck-Institut f\"ur Radioastronomie, Auf dem H\"ugel 69, D-53121 Bonn, Germany}

\author{R.~Karuppusamy}
\affil{Max-Planck-Institut f\"ur Radioastronomie, Auf dem H\"ugel 69, D-53121 Bonn, Germany}

\author{D.~H.~F.~M.~Schnitzeler}
\affil{Max-Planck-Institut f\"ur Radioastronomie, Auf dem H\"ugel 69, D-53121 Bonn, Germany}
\affil{Bendenweg 51, 53121 Bonn, Germany}

\author{H.~Falcke}
\affil{Department of Astrophysics, Institute for Mathematics, Astrophysics and Particle Physics,\\ Radboud University, PO Box 9010, 6500 GL Nijmegen, The Netherlands}
\affil{ASTRON, PO Box 2, 7990 AA Dwingeloo, The Netherlands}
\affil{Max-Planck-Institut f\"ur Radioastronomie, Auf dem H\"ugel 69, D-53121 Bonn, Germany}

\author{M.~Kramer}
\affil{Max-Planck-Institut f\"ur Radioastronomie, Auf dem H\"ugel 69, D-53121 Bonn, Germany}
\affil{Jodrell Bank Centre for Astrophysics, School of Physics and Astronomy, The University of Manchester, Manchester M13 9PL, UK}

\author{L.~G.~Spitler}
\affil{Max-Planck-Institut f\"ur Radioastronomie, Auf dem H\"ugel 69, D-53121 Bonn, Germany}

\author{P.~Torne}
\affil{Instituto de Radioastronom\'ia Milimetrica, Avda. Divina Pastora 7, N\'ucleo Central, E-18012 Granada, Spain}

\author{K.~Liu}
\affil{Max-Planck-Institut f\"ur Radioastronomie, Auf dem H\"ugel 69, D-53121 Bonn, Germany}

\author{G.~C.~Bower}
\affil{Academia Sinica Institute of Astronomy and Astrophysics, 645 N. A'ohoku Place, Hilo, HI 96720, USA}

\author{I.~Cognard}
\affil{Laboratoire de Physique et Chimie de l'Environnement et de l'Espace LPC2E CNRS-Universit{\'e} d'Orl{\'e}ans, F-45071 Orl{\'e}ans, France}
\affil{Station de radioastronomie de Nan{\c c}ay, Observatoire de Paris, CNRS/INSU F-18330 Nan{\c c}ay, France}

\author{A.~G.~Lyne}
\affil{Jodrell Bank Centre for Astrophysics, School of Physics and Astronomy, The University of Manchester, Manchester M13 9PL, UK}

\author{B.~W.~Stappers}
\affil{Jodrell Bank Centre for Astrophysics, School of Physics and Astronomy, The University of Manchester, Manchester M13 9PL, UK}

\begin{abstract}
Polarised radio emission from \magnt{} has already been used to
investigate the strength of the magnetic field in the Galactic Centre,
close to Sagittarius~A*. Here we report how persistent radio emission
from this magnetar, for over four years since its discovery, has
revealed large changes in the observed Faraday rotation measure, by up
to 3500\,rad\,m$^{-2}$ (a five per cent fractional change). From
simultaneous analysis of the dispersion measure, we determine that
these fluctuations are dominated by variations in the projected
magnetic field, rather than the integrated free electron density,
along the changing line of sight to the rapidly moving magnetar. From
a structure function analysis of rotation measure variations, and a
recent epoch of rapid change of rotation measure, we determine a
minimum scale of magnetic fluctuations of size $\sim 2$\,au at the
Galactic Centre distance, inferring \magnt{} is just $\sim 0.1$\,pc
behind an additional scattering screen.
\end{abstract}


\section{Introduction}
Measurements of Faraday Rotation in the polarised emission of radio
sources can be used to examine the strength and structure of the
magnetic field in the interstellar medium \citep{bw+13}.  A recent
notable example is the radio-loud magnetar, \magnt{}, which displays a
high rotation measure (RM) of $-66960\pm50$\,rad\,m$^{-2}$, second
only in the Galaxy to the
$\text{RM}=(-4.3\pm0.1)\times10^5$\,rad\,m$^{-2}$ of the supermassive
black hole candidate, Sagittarius~A$^*$ (Sgr~A$^*$), caused
predominantly by the accretion flow on scales smaller than the
Bondi-Hoyle radius \citep{bwf+03}.  This magnetar therefore allowed
first-order estimates of the strength of the magnetic field at the
beginnings of the Bondi-Hoyle accretion radius of \sgra{}; $\sim
8$\,mG at scales of $\sim 0.1$\,pc \citep{efk+13}.  The magnitude and
spatial or time variability of the magnetic field also allows models
of the accretion flow to be investigated \citep{2011MNRAS.415.1228P}.

\magnt{} has also been used to examine the scattering of radio waves
towards the Galactic Centre (GC) by measurements of both the temporal
pulse broadening \citep{sle+14} and the angular image broadening
\citep{bdd+14}. Combination of the two measurements
indicates the principal scattering screen toward the GC is
$5.8\pm0.3\,\text{kpc}$ in front of the magnetar \citep{bdd+14}.

Atypically for magnetars, \magnt{} has remained active in the radio
band for over four years since its discovery at X-ray wavelengths in
2013 \citep{kbk+13, mgz+13}. This has allowed repeated measurements of
the RM and dispersion measure (DM). Because the magnetar has a total
proper motion of $6.37\pm0.16$\,mas\,yr$^{-1}$ \citep{bdd+15}, the
time-variations in the measurements of DM and RM presented here occur
along different sightlines. The physical scales probed in the GC are
therefore directly related to the observing cadence, and the
transverse velocity of the magnetar.

In this paper, long term polarimetric observations of \magnt{} with the
Effelsberg and Nan\c cay radio telescopes are presented. Section 2 gives
a description of the observational campaign and data analysis
techniques used. In Section 3 the results of the analysis are
presented, and in Section 4 we turn to physical interpretations of the
results.

\section{Observations}
Following the discovery of radio pulsations from \magnt{} in April 2013,
this source has been monitored with three European radio telescopes
operating at complementary observing frequencies. These are the
Effelsberg radio telescope, the Nan\c cay Radio Telescope (NRT) and
the Jodrell Bank, Lovell radio telescope. In this work we only refer
to measurements from the Effelsberg telescope and the NRT because
observations with the Lovell telescope, at a lower frequency of
1.4\,GHz, suffer from instrumental depolarisation due to the large RM.

\subsection{Effelsberg}
\magnt{} is observed with the Effelsberg telescope typically on a
monthly basis, and fortnightly since January 2017.  One-hour
observations at central frequencies of 8.35\,GHz and 4.85\,GHz are
recorded with the PSRIX backend \citep{lkg+16}.

The 500\,MHz bandwidth provided by the PSRIX backend is first
digitised and split into 512 or 1024 channels when observing at 8.35\,GHz
and 4.85\,GHz, respectively.  The channelised data are then
dedispersed at a DM of 1778\,pc\,cm$^{-3}$, the initial value DM
measured by \citet{efk+13} and finally folded at the period of the
magnetar to create `single pulse profiles'.

Since March 2017 the new `C+' broadband receiver is available for
pulsar observations. Two 2\,GHz bands (covering 4 to 8\,GHz) are fed
into the new PSRIX2 backend, consisting of two
CASPER\footnote{https://casper.berkeley.edu/} ROACH2 boards. Each
board digitises the signal and acts as a full-Stokes spectrometer,
creating 2048 frequency channels every 8\,$\mu$s. The data are later
dedispersed and folded to create single pulses profiles.  During the
commissioning phase of PSRIX2, the C+ observations replace the
8.35\,GHz observations.
 
\subsection{Nan\c cay}
Observations with the NRT were carried out with the NUPPI
instrumentation on average every 4 days between May 2013 and August
2014, before resuming at a monthly cadence between January and July
2017.  The setup of the observations was already presented in
\citet{efk+13,sle+14,tek+15}. We briefly summarize it here.  A
bandwidth of 512\,MHz centred at 2.5\,GHz is split into 1024 channels
and coherently dedispersed using an initial DM value of
1840\,pc\,cm$^{-3}$ (the best DM value derived from timing of the
scattered pulse profiles) then folded at the period of the magnetar
and written to disk every 30\,s.

\subsection{Post-processing and calibration}
All the data presented here are corrected for the gain and phase
difference between the feeds of the various receivers used. This is
achieved by standard pulsar calibration techniques that use
observations of a polarised pulsed noise diode.  Large $|$RM$|$s, in
combination with wide frequency channels, lead to instrumental
depolarization. Following the analysis presented in \cite{sl2015}, we
calculate that an $|$RM$|$ of $7\times 10^4$\,rad\,m$^{-2}$
depolarizes the signal by only 4 per cent in the 2.5\,GHz band; at the
higher observing frequencies, this effect is negligible.  The data
reduction is achieved with the standard \textsc{PSRCHIVE} package
\citep{hvm04}.

\section{Results}
Monitoring of the dispersive and polarisation properties of \magnt{}
over a period of $\sim 54$ months has revealed a rather constant DM of
$1762\pm11$\,cm\,pc$^{-3}$, while variations in RM
$>3500$\,rad\,m$^{-2}$ are observed. Because RM is proportional to
both projected magnetic field strength $B_{\parallellos}$, and the
free electron density $n_\textrm{e}$, along the line-of-sight $s$,
${\rm RM} \propto \int B_{\parallellos} (s)
n_\textrm{e}(s) \text{d}s$, it is important to disentangle these
quantities to understand the physical mechanisms causing the
variations in RM. For \magnt{}, which has the highest DM amongst all
the known pulsars, measurement of the DM is influenced by pulse
scattering. In this section the methods used to measure both the DM
and RM are described, as well as the results of our monitoring
campaign.

\subsection{Dispersion Measure variations}
\label{ref:dm}
To accurately measure the DM and remove the bias caused by the
scattering of the pulse profile (and possibly the variations of it),
we modelled both scattering and DM simultaneously over a range of
frequency subbands for each NUPPI, PSRIX and PSRIX2 observations.
Given the low amount of dispersion across the band of the PSRIX data
at 8.35\,GHz ($\sim 12$\,ms), we did not apply this technique to
these data.

Following \citet{sle+14}, we use scattered Gaussian pulse function to
model the single pulses observed between 4 and 8\,GHz and the averaged
pulse profile at 2.5\,GHz.  However, in contrast to \citet{sle+14}, we
did not correct for the jittering of the single pulses to create an
average `de-jittered' profile. Instead, we model simultaneously some
of the brightest single pulses in each observation with different
Gaussian widths $\sigma$ to increase the significance of our results.

The scattered Gaussian pulse profile function for a single channel is
given by Equation~3 of \citet{sle+14}. We extend it here for multiple
channels to include the DM as a parameter.  We can therefore write the
likelihood $\Lambda$ as
\begin{equation}
\log \Lambda = \frac{1}{2} \sum_{i}^{N_\text{p}}
\sum_{j}^{N_\text{c}}\sum_{k}^{N_\text{b}} \left\{P_{ijk} -
T_{ijk}\right\}+ \text{const},
\end{equation}
where $P_{ijk}$ is the observed single pulse profile $i$ with
frequency channel $j$ and profile bin $k$ over $N_\text{p}$ 
profiles included in the modelling with $N_\text{c}$
frequency channels and $N_\text{b}$ phase bins. $T_{ijk}$ is
the modelled scattered profile,
\begin{multline}
T_{ijk} = \frac{A_{ij}\sqrt{\pi}} {2} \sigma_{i}
e^{\frac{\sigma_{i}^{2} - 4\phi_{ijk}\tau} {4\tau^{2}}} \\ \times
\left\{ 1 + \text{erf} \left( \frac{2\tau(\phi_{ijk}-\phi_{0,ij}) -
  \sigma_i^2} {2\tau\sigma_i} \right) \right\} + b_{ij}.
\end{multline}
$A_{ij}$ and $b_{ij}$ are, respectively, the amplitude and the
baseline offset of channel $j$ from profile $i$. $\phi_{0,ij}$ is the
phase centre of the Gaussian of width $\sigma_i$, but delayed in each
channel because of dispersion. $\phi_{0,ij}$ is given by $\phi_{0,ij}
= \phi_{0,i} + k \text{DM} \times (1/f_{i}^{2} -1/f_{0}^2)$, where the
dispersion constant $k=4.15\times10^3$\,MHz$^2$\,cm$^3$\,pc$^{-1}$\,s.
Finally, $\tau$ is the scattering time at a reference frequency $f_0$.
The dimensionality of the model is therefore $2 + N_\text{p} \times (
2+ 2\times N_\text{c}) $, where $N_\text{p}$ is 1 for the NRT averaged
profile, typically 2 to 4 for the Effelsberg single pulse observations
and $N_\text{c}$ range from 8 to 16.

\begin{figure}
\includegraphics[height=\columnwidth,angle=-90]{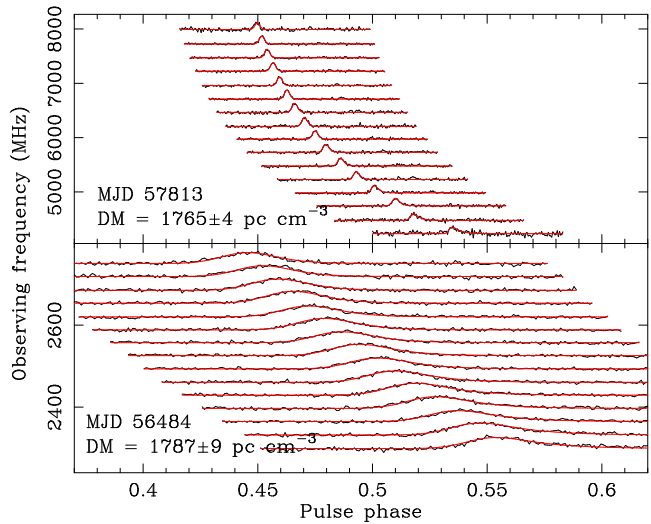}
\caption{Scattered and dispersed observed pulse profiles of \magnt{}
  (over a selected phase window) are shown in black color for a set of
  16 subbands from a C+ single pulse profile (top panel) and a
  2.5\,GHz daily averaged profile (bottom panel). The red curves show
  our model of scattered and dispersed Gaussian profiles taken from
  the maximum likelihood solution. The MJD of the observation and the
  estimated DM are indicated at the bottom-left part of each panel.}
\label{fig:scattering}
\end{figure}

Given the large dimensionality of the model, we implemented the nested
sampling tool \textsc{POLYCHORD} \citep{hhl15b} to efficiently sample
the parameter space instead of using traditional $\chi^2$-fitting
techniques. The priors on all parameters are set to be uniform, except
for $A_{ij}$ where we use log-uniform priors. We refer to one-sigma
error bars as the 68.3 per cent contours of the one-dimensional
marginalized posterior distribution of each parameter. The resulting
code can be found
online\footnote{https://github.com/gdesvignes/scattering}.  An example
of our modelling results can be seen in Figure~\ref{fig:scattering}.

\begin{figure}
\includegraphics[height=\columnwidth,angle=-90]{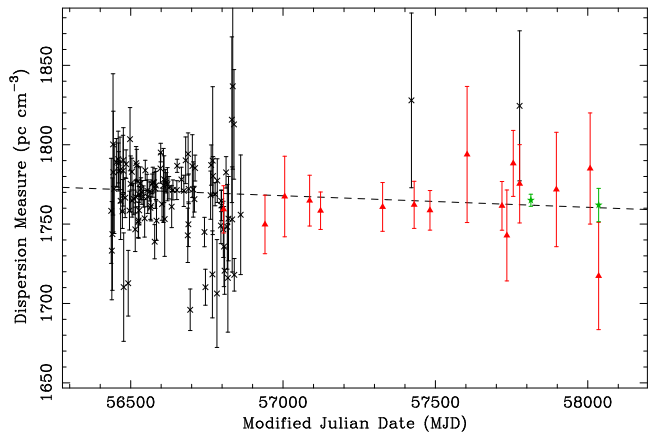}
\caption{DM variations of the GC magnetar. The black crosses, red
  triangles and green stars show the DM values (with one-sigma error bars) from the
  modelling of the scattered profiles from the 2.5\,GHz, 4.85\,GHz and C+ data,
  respectively. The dashed line represents the linear least-squares
  fit to the data. The MJD and modelled DM are reported in the
  bottom-left part of each panel.}
\label{fig:DM}
\end{figure}

The DM results of the modelling of the Effelsberg and NRT data are
shown in Figure~\ref{fig:DM}. A linear least-squares fit to the DM
values shows a marginal decrease of approximately 0.6 per cent over
four years (two-sigma consistent with no DM change) with
DM=$1762\pm11$\,cm\,pc$^{-3}$ as measured in October 2017. Results
from the scattering measurements that come out of this joint analysis
will be published elsewhere.

\subsection{Rotation Measure variations}
\label{ref:rm}
\begin{figure}
\includegraphics[height=\columnwidth,angle=-90]{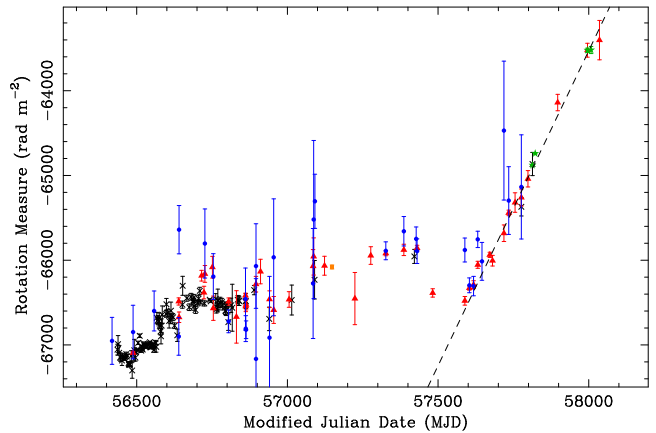}
\caption{Variation of RM as a function of MJD. The black crosses, red
  triangles, blue points and green stars show the RM values estimated
  from the 2.5\,GHz, 4.85\,GHz, 8.35\,GHz and C+ data,
  respectively. The orange rectangle shows the value from the ATCA
  data recorded at 5.4\,GHz \citep{sef+16}. The black dashed line
  shows the fitted slope of the recent epoch of large gradient in RM
  variation.}
\label{fig:RM}
\end{figure}

To determine the Faraday rotation of the linearly polarised emission
of the magnetar, we used the RM synthesis method \citep{bd05} and
fitted the wrapping of the Stokes vector Q and U under the pulse
window as previously done in \citet{efk+13}. We employed this
technique to extract the RM for all our data. In contrast to our DM
results, Figure~\ref{fig:RM} shows the rapid and non-monotonic
evolution of the RM during our 4 years of observations. Since the
initial $\text{RM}=-66960\pm50$\,rad\,m$^{-2}$ reported in
\citet{efk+13}, the RM changed by over 3500 units to
$\text{RM}=-63402\pm232$\,rad\,m$^{-2}$ in October 2017 (a relative
change of 5.3 per cent). Our measurements are also consistent with the
$\text{RM}=-66080\pm24$\,rad\,m$^{-2}$ reported by \citet{sef+16}
using Australian Telescope Compact Array (ATCA) observations at
5.4\,GHz. After a steady increase in the first few months, the RM
increased abruptly around MJD 56566 followed by another steady
increase until MJD 57450. Then the RM diminished up to MJD 57600 before
increasing over the last year of observations.  A least-squares fit to
the data collected over the last year shows that RM changed by about
$7.4$\,rad\,m$^{-2}$ per day.  It is worth noting that two-weeks
of follow-up observations of the magnetar SGR~1806$-$20 after an outburst
phase did not show evidence for RM variations \citep{gkg+05}.

\subsection{Polarisation fraction}
\begin{figure}
\includegraphics[height=\columnwidth,angle=-90]{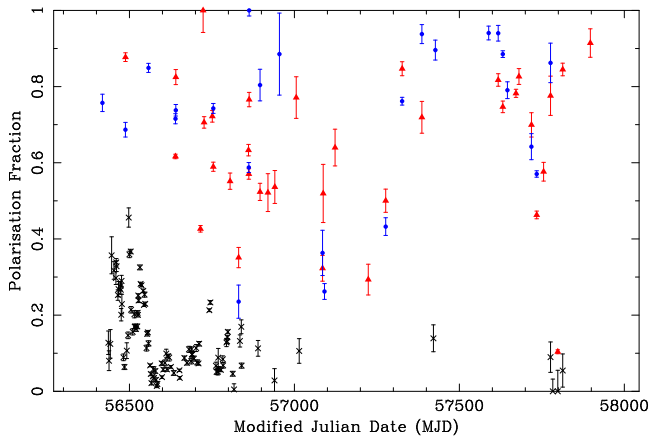}
\caption{Polarisation fraction as a function of time. The black
  crosses, red triangles and blue points show the values from the
  2.5\,GHz, 4.85\,GHz and 8.35\,GHz data, respectively. The
  polarisation fraction at 2.5\,GHz is consistently lower than at the
  higher frequencies.}
\label{fig:pol_fraction}
\end{figure}

Figure~\ref{fig:pol_fraction} shows the evolution of the linear
polarisation fraction \lf{} as a function of time for all our
data. These results show large variations in \lf{} at all frequencies,
with \lf{} at 2.5\,GHz consistently lower than at higher
frequencies, by a factor of between 2 and 10.

\section{Discussion}
The observations presented here show a clear dichotomy between the
properties of the DM and RM variations toward PSR~J1745$-$2900; the DM
varies marginally while the RM shows very large variations. This is
not entirely unexpected since it has been suggested that the RM is
likely caused by local magnetic phenomena, somewhere in the last few
parsecs toward the GC, whereas most of the DM is accumulated along the
entire line of sight to the magnetar \citep{efk+13}.  We note the RM
variations observed are three orders of magnitude above what might be
expected from ionospheric contributions \citep{ymh+11}.

\subsection{Structure function analysis}
\label{sec:sf}
\begin{figure}
\includegraphics[height=\columnwidth,angle=-90]{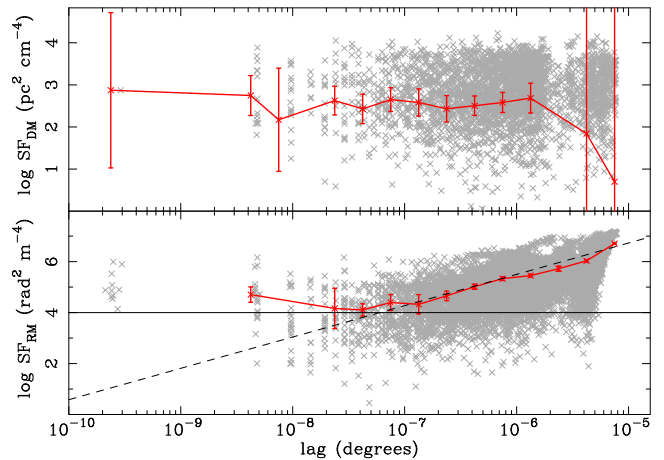}
\caption{Structure functions, corrected for measurement errors, for
  the DM (top panel) and RM (bottom panel). For clarity, the error
  bars for each lag (grey crosses) are not shown. The red points
  represent the averaged values over intervals of 0.25 in log
  scale. The horizontal black line delimits RM variations of
  100\,rad\,m$^{-2}$. The black dashed line shows the linear fit to
  the average values of ${\rm SF}_{\rm RM}$ allowing to give an
  estimate on the angular scale at which the RM variations occur.}
\label{fig:struct_function}
\end{figure}

As already mentioned, \magnt{} has a proper motion \citep{bdd+15}, so
our measurements of DM and RM can be used to probe the angular and
physical scales on which these fluctuations occur. For sparsely and
irregularly sampled data, like these, power spectrum analyses are
unsuitable. A frequently used alternative is the structure function of
variations, see e.g. \citet{ms+96}. Following their notation, the
ensemble average second order structure function in DM and RM can be
written ${\rm SF}_{\rm DM} = \langle [{\rm DM}(\theta)-{\rm
    DM}({\theta}+\delta{\theta})]^2 \rangle$ and ${\rm SF}_{\rm RM} =
\langle [{\rm RM}(\theta)-{\rm RM}({\theta}+\delta{\theta})]^2
\rangle$, respectively, where ${\theta}$ is any given line of sight
and $\delta\theta$ is the angular separation or `lag' of two
measurements. In this work the lag is given by the total magnitude of
the proper motion (6.37\,mas\,yr$^{-1}$) multiplied by the separation
in time of two DM or RM measurements.  In addition, we corrected each
pair of DM and RM used in the structure function for noise bias, by
subtracting ($\delta \textrm{DM}(\theta)^2+\delta
\textrm{DM}({\theta}+\delta{\theta})^2$) and ($\delta
\textrm{RM}(\theta)^2+\delta \textrm{RM}({\theta}+\delta{\theta})^2$),
respectively.  In Figure \ref{fig:struct_function} we present the
results from the structure function analysis. In the top panel,
averaged values of ${\rm SF}_{\rm DM}$ are unchanging across all
angular scales from ${10^{-10}}$\,deg to ${10^{-5}}$\,deg. This is
representative of the flat fit of the DM time series in
Section~\ref{ref:dm}. The bottom panel of Figure
\ref{fig:struct_function} shows ${\rm SF}_{\rm RM}$ over the same
interval in angular scale. Unlike ${\rm SF}_{\rm DM}$, ${\rm SF}_{\rm
  RM}$ begins rising at angular scales over ${10^{-8}}$\,deg,
equivalent to a physical size of a few au at a GC distance of
$8.3$\,kpc \citep{gef+09}. We define this size as the approximate
smallest-scale at which magnetic structure in the GC is observed in
our data. 

The fitted slope of ${\rm SF}_{\rm RM}$ (black dashed line), $l=1.23\pm0.13$, is potentially indicative of Kolmogorov turbulence (where $l=5/3$) assuming isotropic and stationary fluctuations in RM \citep{scs+84,lsc+90,hgm+04}.

\subsection{A physical model based on the depolarisation}
\label{sec:depol}
Figure~\ref{fig:pol_fraction} shows that PSR~J1745$-$2900 depolarises
rapidly with decreasing frequency.  This could be intrinsic to the
magnetar, or caused by a propagation effect. Because the former has
not been observed in magnetars \citep[e.g.][]{ksj+07}, in this section
we model depolarisation due to propagation through the interstellar
medium.

For this purpose we invoke a secondary scattering screen, close to the
magnetar in the GC, in addition to the distant screen ($\sim6$\,kpc
from the GC) responsible for the temporal scatter broadening
\citep{bdd+14, 2014evn..confE..66W}. The distant screen cannot produce
the rapid RM variations we observe due to its large size (160\,mas at
2.5\,GHz) because spiral arms do not host strong magnetic fields
\citep{hav15}. Scattering screens located close to the GC (<700\,pc) have
been previously indicated for other GC pulsars \citep{ddb+17}.  As the
magnetar moves behind or through a magnetized, ionized gas, the
variation in RM that we observe over time also imprints itself as a
gradient in RM across this secondary scattering disk.  Following
\citet{sbl2015}, we can show that RM variations of order
100\,rad\,m$^{-2}$ within a scattering disk (Gaussian disk with a FWHM
of 0.2-0.5\,mas or a uniformly lit circular disk with diameter
$<$~1\,mas at 2.5\,GHz) is enough to explain little or no
depolarization at the highest observing frequencies, and the strong or
even complete depolarization at 2.5\,GHz.  At this frequency an RM
gradient of $100$\,rad\,m$^{-2}$ across the scattering disk would lead
to PA differences up to $\sim 80$\,deg, leading to strong
depolarisation.

By considering both measurements from the RM structure function
analysis outlined in Section~\ref{sec:sf}, and the recently measured
large gradient in RM (the fitted slope described in
Section~\ref{ref:rm}), fluctuations of order $100$\,rad\,m$^{-2}$ are
readily observable in our data set and occur at angular-scales, or
time-scales, of the order tenths of milliarcseconds and weeks
respectively.

The solid horizontal black line at ${\rm SF}_{\rm RM}=10000\,{\rm
  rad^{2}\,m^{-4}}$ in Figure~\ref{fig:struct_function} defines the
limit where changes in RM of 100\,rad\,m$^{-2}$ occur. The intercept
of this line with the fitted averages of ${\rm SF}_{\rm RM}$ gives the
approximate angular scale at which these fluctuations begin to occur;
which is around $6\times 10^{-8}$\,deg ($\sim 0.22$\,mas or a physical
scale of $\sim 1.8$~au at a GC distance of 8.3\,kpc). The fitted slope
in RM in Figure~\ref{fig:RM} indicates a change in RM of
100\,rad\,m$^{-2}$ occurs on time-scales of just two weeks (angular
displacement of $\sim 0.24$\,mas, physical scale of $\sim2.0$\,au at
distance of 8.3\,kpc). Much shorter term variations ($<1$\,hr) have
been ruled out. The average value of these two measured physical
scales maps to an upper bound on the scattering disk size $b$ of $\sim
1.9$\,au.

The geometric scattering time delay, $\tau$, depends upon both the
magnetar's distance to the screen, $z$, and its size, $b$, through
$\tau=b^2/2cz$. $\tau$ is bounded by the observed scattering time at
2.5\,GHz, and is likely substantially shorter since most of the
scattering delay occurs at the midway screen
\citep{2014evn..confE..66W}. Following \citet{sle+14} we use
$\tau<0.04$\,s giving $z\gtrsim 0.1$\,pc. Remarkably, this places the
secondary scattering screen at least 0.1\,\text{pc} in front of
\magnt{}, the same distance as the projected offset from \sgra;
therefore placing the screen close to the Bondi-Hoyle accretion
radius.  The observed fluctuations in magnetic field are thus
potentially occurring within the black hole reservoir.  The two main
assumptions in this model are the isotropy of the RM variations, and
the attribution of depolarisation to propagation effects.

The constancy in electron density toward \magnt{}, indicated by our DM
measurements, therefore suggests an additional scattering screen local
to the GC and embedded in a strong ambient magnetic field \citep[$\sim
  8$\,mG;][]{efk+13}.  Fluctuations in this field, 
 of $\sim 12\,\mu$G
starting at sizes of $\sim 2$\,au and going up to $\sim 400\,\mu$G on
the largest measured scales of $\sim 300$\,au occur either due to
changes in its strength or orientation. Our data also suggest the RM fluctuations
are potentially Kolmogorov in nature.

Future observations of
\magnt{} will better constrain the magnetic field structure around
\sgra{} and can potentially be used to test if the accretion flow
becomes magnetically dominated \citep{2003ApJ...596L.207P}.

\acknowledgments
The authors acknowledge financial support by the European Research
Council for the ERC Synergy Grant BlackHoleCam under contract
no. 610058.  KJL received support from 973 program 2015CB857101, NSFC
U15311243, XDB23010200, 11690024, and funding from Max-Planck Partner
Group. This work was based on observations with the 100-m telescope of
the Max-Planck-Institut f\"ur Radioastronomie at Effelsberg. The Nan\c
cay radio Observatory is operated by the Paris Observatory, associated
to the French Centre National de la Recherche Scientifique (CNRS) and
to the Universit\'e d'Orl\'eans. The authors thank D.D.~Xu, R.~Beck
and O.~Wucknitz and M.~Haverkorn for helpful discussions.

\end{document}